\documentclass[aps,pra,superscriptaddress,floatfix,showpacs,11pt, column,showkeys]{revtex4}
\usepackage{bbm}
\usepackage{amsfonts}
\usepackage[dvips]{graphicx}
\usepackage{amsmath,amsfonts,amssymb,graphics,graphicx,epsfig,color,times,bbm}

\begin{document}

\title{Nonlinear steering criteria for arbitrary two-qubit quantum systems}

\author{Guo-Zhu Pan\footnote{panguozhu@wxc.edu.cn}}
\affiliation{School of Electrical and Photoelectric Engineering, West Anhui University, Lu'an, 237012, China}

\author{Ming Yang\footnote{mingyang@ahu.edu.cn}}
\affiliation{School of Physics {\&} Materials Science, Anhui University, Hefei 230601, China}

\author{Hao Yuan}
\affiliation{School of Electrical and Photoelectric Engineering,
West Anhui University, Lu'an, 237012, China}
\affiliation{School of Physics {\&} Materials Science, Anhui University, Hefei 230601, China}

\author{Gang Zhang}
\affiliation{School of Electrical and Photoelectric Engineering, West Anhui University, Lu'an, 237012, China}

\author{Jun-Long Zhao}
\affiliation{School of Physics {\&} Materials Science, Anhui University, Hefei 230601, China}
\begin{abstract}
\textbf{ Abstract:} By employing Pauli measurements, we present some nonlinear steering criteria applicable for arbitrary two-qubit quantum systems and optimized ones for symmetric quantum states. These criteria provide sufficient conditions to witness steering, which can recover the previous elegant results for some well-known states. Compared with the existing linear steering criterion and entropic criterion, ours can certify more steerable states without selecting measurement settings or correlation weights, which can also be used to verify entanglement as all steerable quantum states are entangled.
  \end{abstract}

\maketitle

\section{Introduction}

Quantum steering describes the ability of one observer to nonlocally
affect the other observer's state through local measurements, which
was first noted by Einstein, Podolsky and Rosen (EPR) for arguing
the completeness of quantum mechanics in 1935 \cite{ein}, and later
introduced by Schr\"{o}dinger in response to the well-known EPR
paradox\cite{sch}. After being formalized by Wiseman et al. with a
local hidden variable (LHV)-local hidden state model in 2007
\cite{wis}, quantum steering has attracted increasing attention and
been explored widely. Steerable states were shown to be advantageous
for tasks involving secure quantum teleportation \cite{rei, ros},
quantum secret sharing \cite{walk, kog}, one-sided
device-independent quantum key distribution \cite{bra} and channel
discrimination \cite{pia}.

Quantum steering is one form of quantum correlations intermediate between quantum entanglement \cite{horo} and Bell nonlocality \cite{bell}. It has been demonstrated that a quantum state which is Bell nonlocal must be steerable, and a quantum state which is steerable must be entangled \cite{jone, brun}. One distinct feature of quantum steering which differs from entanglement and Bell nonlocality is asymmetry. That is, there exists the case when Alice can steer Bob's state but Bob cannot steer Alice's state, which is referred to as one-way steerable and has been demonstrated in theory \cite{bow} and experiment \cite{han, wol}.

Quantum steering is the failure description of the local hidden variable-local hidden state models to reproduce the correlation between two subsystems, which can be witnessed by quantum steering criteria. Recently, a lot of steering criteria have been developed to distinguish steerable quantum states from unsteerable ones. In Ref. \cite{sau}, the linear steering criteria was introduced for qubit states. In Ref. \cite{sch2}, the steering criteria from entropic uncertainty relations were derived, which can be applicable for both discrete and continuous variable systems. Subsequently, the steering criteria via covariance matrices of local observables \cite{ji} and local uncertainty relations \cite{zhen} in arbitrary-dimensional quantum systems were presented. Recently, Refs. \cite{zhe1, zhe2} generalized the linear steering criteria to high-dimensional systems. Although these criteria work well for a number of quantum states, most of them require constructing appropriate measurement settings or correlation weights in practice, which increases the complexities of the detecting inevitably. The development of the universal criterion to detect steering is still one vexed question.

In this paper, we first present some steering criteria applicable for arbitrary two-qubit quantum systems, then optimize them for symmetric quantum states, and finally we provide a broad class of explicit examples including two-qubit Werner states, Bell diagonal states, and Gisin states. Compared with the existing linear
steering criterion and entropic criterion, ours can certify more
steerable states without selecting measurement settings or
correlation weights, which can also be used to verify entanglement
as all steerable quantum states are entangled.

\section{Nonlinear steering criteria for arbitrary two-qubit quantum systems}
Suppose two separate parties, Alice and Bob, share a two-qubit
quantum state on a composite Hilbert space
$\mathcal{H}=\mathcal{H}_{A}\otimes\mathcal{H}_{B}$. The steering is
defined by the failure description of all possible local hidden variable-local hidden state models in
the form \cite{wis, jone}
\begin{equation}\label{1}
P(a,b|A,B;W)=\sum_{\lambda}P(a|A;\lambda)P(b|B;\rho_{\lambda})p_{\lambda},
\end{equation}
where $P(a,b|A,B;W)$ are joint probabilities for Alice and Bob's
measurements $A$ and $B$, with the results $a$ and $b$,
respectively; $p_{\lambda}$ and $P(a|A;\lambda)$ denote some
probability distributions involving the LHV $\lambda$, and
$P(b|B;\rho_{\lambda})$ denotes the quantum probability of outcome
$b$ given measurement $B$ on state $\rho_{\lambda}$. $W$ represents
the bipartite state under consideration. In other words, a quantum
state will be steerable if it does not satisfy Eq.(1). Within the
formulation, we propose a nonlinear steering criterion that can be
used to certify a wide range of steerable quantum states for two-qubit quantum systems.

\emph{Theorem 1.} If a given two-qubit quantum state is unsteerable from Alice to Bob (or Bob to Alice), the following inequality holds:
\begin{equation}
\sum\limits_{i=1}\limits^{3}\sum\limits_{j=1}\limits^{3}\langle\sigma_{i}\otimes\sigma_{j}\rangle^{2}\leq1,
\end{equation}
where $\sigma_{i,j}$ ($i,j=1,2,3$) are Pauli operators.

\emph{Proof.} Suppose Alice and Bob share a two-qubit quantum state $\rho_{AB}$ on a composite Hilbert space, both of them perform $N$ measurements on their own states, which are denoted by $A_{k}$ and $B_{l}$, respectively. Here $B_{l}$ is a quantum observable while $A_{k}$ have no such constraint, $k (l)$ ($k (l)=1,2,\cdots,N$) labels the \emph{k}th (\emph{l}th) measurement setting for Alice (Bob). If the state is unsteerable from Alice to Bob, we have the following inequality
\begin{eqnarray}
 \nonumber
&&\sum\limits_{k=1}\limits^{N}\sum\limits_{l=1}\limits^{N}\langle A_{k}\otimes B_{l}\rangle^{2}\\ \nonumber
&=& \sum\limits_{k=1}\limits^{N}\sum\limits_{l=1}\limits^{N}\left(\sum\limits_{a_{k},b_{l}}a_{k}b_{l}P(a_{k},b_{l}|A_{k},B_{l};\rho_{AB})\right)^2\\ \nonumber
&\leq&\sum\limits_{\lambda}\left(p_{\lambda}\sum\limits_{k=1}\limits^{N}\left[\sum\limits_{a_{k}}a_{k}P(a_{k}|A_{k},\lambda)\right]^2\sum\limits_{l=1}\limits^{N}\left[\sum\limits_{b_{l}}b_{l}P(b_{l}|B_{l},\rho_{\lambda})\right]^2\right)\\ \nonumber
&=&\sum\limits_{\lambda}p_{\lambda}\left(\sum\limits_{k=1}\limits^{N}\langle A_{k}\rangle^{2}_{\lambda}\sum\limits_{l=1}\limits^{N}\langle B_{l}\rangle^{2}_{\rho_{\lambda}}\right)\\ \nonumber
&\leq&\eta \sum\limits_{\lambda}p_{\lambda}\left(\sum\limits_{k=1}\limits^{N}\langle A_{k}^{2}\rangle_{\lambda}\right)\max\limits_{\{\rho_{\lambda}\}}\left(\sum\limits_{l=1}\limits^{N}\langle B_{l}\rangle^{2}_{\rho_{\lambda}}\right)\\
&=&\eta\sum\limits_{k=1}\limits^{N}\langle A_{k}^{2}\rangle
C_{B}=\eta C_{A}C_{B},
\end{eqnarray}
where $C_{A}=\sum\limits_{k=1}\limits^{N}\langle
A_{k}^{2}\rangle,C_{B}=\max\limits_{\{\rho_{\lambda}\}}\left(\sum\limits_{l=1}\limits^{N}\langle
B_{l}\rangle^{2}_{\rho_{\lambda}}\right)$. The parameter $\eta$
($0\leq\eta\leq1$) is a constant, which is used to adjust the value
to the appropriate bound. The first inequality follows from the fact $\sum_{k=1}^{N}\sum_{l=1}^{N}(\alpha_{k}\beta_{l})^{2}\leq\sum_{k=1}^{N}\alpha_{k}^{2}\sum_{l=1}^{N}\beta_{l}^{2}$. The second inequality follows from the definition
of $C_{B}$ and the fact $\langle
A_{k}^{2}\rangle_{\lambda}\geq\langle A_{k}\rangle_{\lambda}^{2}$.
If the observables $A_{k}$ and $B_{l}$ are restricted to Pauli
matrices, i.e., $A_{k} (B_{l})=\{\sigma_{1}, \sigma_{2},
\sigma_{3}\}$, one has straightforwardly $C_{A}=3$ and $C_{B}=1$, so
Eq.(3) reduces to
\begin{equation}
\sum\limits_{i=1}\limits^{3}\sum\limits_{j=1}\limits^{3}\langle\sigma_{i}\otimes\sigma_{j}\rangle^{2}\leq\eta'.
\end{equation}
where $\eta'=3\eta$.

As we know, quantum entanglement, quantum steering, and Bell nonlocality are equivalent in the case of pure states \cite{wis, jone, ysx}. For an arbitrary quantum steering criterion, it is preferable to be a sufficient and necessary condition to detect pure states \cite{zhe1, zhe2, zhen}. In order to obtain the optimal value of the parameter $\eta'$, we introduce the pure states as reference states. For any two-qubit state, it can be expressed as
\begin{equation}\label{9}
\rho_{AB}=\frac{1}{4}(\mathbb{I}+\sum_{i=1}^{3}c_{i0}\sigma_{i}\otimes\mathbb{I}+\sum_{j=1}^{3}c_{0j}\mathbb{I}\otimes\sigma_{j}+\sum_{i=1}^{3}\sum_{j=1}^{3}c_{ij}\sigma_{i}\otimes\sigma_{j}),
\end{equation}
where $|c_{ij}|\leq1$ for $i,j=0,1,2,3$. For arbitrary pure states
$\rho_{AB}$, one has straightforwardly
$\sum_{i=1}^{3}c_{i0}^{2}+\sum_{j=1}^{3}c_{0j}^{2}+\sum_{i=1}^{3}\sum_{j=1}^{3}c_{ij}^{2}=3$
due to the fact $tr(\rho_{AB}^{2})=1$. Next we consider two cases, one is that $\rho_{AB}$ be pure separable states, then one achieves
$\sum_{i=1}^{3}\langle\sigma_{i}\otimes\mathbb{I}\rangle^{2}=\sum_{i=1}^{3}c_{i0}^{2}=1,
\sum_{i=1}^{3}\langle\mathbb{I}\otimes
\sigma_{j}\rangle^{2}=\sum_{i=1}^{3}c_{0j}^{2}=1$, and
$\sum_{i=1}^{3}\sum_{j=1}^{3}\langle\sigma_{i}\otimes\sigma_{j}\rangle^{2}=\sum_{i=1}^{3}\sum_{j=1}^{3}c_{ij}^{2}=1$,
which result in $\eta'\geq1$ due to the fact that all pure separable
states are unsteerable. The other is that $\rho_{AB}$ be pure entangled states, then one attains
$\sum_{i=1}^{3}\langle\sigma_{i}\otimes\mathbb{I}\rangle^{2}=\sum_{i=1}^{3}c_{i0}^{2}<1,
\sum_{i=1}^{3}\langle\mathbb{I}\otimes
\sigma_{j}\rangle^{2}=\sum_{i=1}^{3}c_{0j}^{2}<1$, and
$\sum_{i=1}^{3}\sum_{j=1}^{3}\langle\sigma_{i}\otimes\sigma_{j}\rangle^{2}=\sum_{i=1}^{3}\sum_{j=1}^{3}c_{ij}^{2}>1$,
which result in $\eta'\leq1$ due to the fact that all pure entangled
states are steerable \cite{zhe1, zhe2, zhen}.  So the optimal value
of the parameter $\eta'=1$. This gives the proof of Theorem 1.

In this way, we derive the steering criterion for arbitrary two-qubit
quantum systems. Whatever strategies Alice and Bob choose, a
violation of inequality (2) would imply steering.

In the following we further develop steering criterion by introducing quantum correlation matrix of local observables. Given a quantum state $\rho$ and observables $\{O_{k}\} (k = 1,2, . . . ,n)$, an $n\times n$ symmetric covariance matrix $\gamma$ is defined as \cite{ji}
\begin{equation}\label{5}
\gamma_{kk'}(\rho)=(\langle O_{k}O_{k'}\rangle+\langle O_{k'}O_{k}\rangle)/2-\langle O_{k}\rangle\langle O_{k'}\rangle.
\end{equation}

Now, let us consider a composite system $\rho_{AB}$ and a set observables $\{O_{m}\}=\{\sigma_{i}\otimes\sigma_{j}\} (i,j=1,2,3, m=3(i-1)+j)$. Similarly, the covariance matrix can be constructed as
\begin{equation}\label{6}
\gamma_{mm'}(\rho_{AB})=(\langle O_{m}O_{m'}\rangle+\langle O_{m'}O_{m}\rangle)/2-\langle O_{m}\rangle\langle O_{m'}\rangle.
\end{equation}

Obviously, the diagonal elements of the covariance matrix stand for the variance of  the observables $\{O_{m}\}$.

\emph{Corollary 1.} If a given quantum state $\rho_{AB}$ is unsteerable, the sum of the eigenvalues of the covariance matrix $\gamma_{mm'}(\rho_{AB})$ of the observables $\{O_{m}\}=\{\sigma_{i}\otimes\sigma_{j}\} (i,j=1,2,3, m=3(i-1)+j)$ must satisfied
\begin{equation}\label{7}
\sum\limits_{k=1}\limits^{9}\lambda_{k}\geq 8,
\end{equation}
where $\lambda_{k}$ is the eigenvalue of the covariance matrix $\gamma_{mm'}(\rho_{AB})$.

\emph{Proof.} For an unsteerable state $\rho_{AB}$, one has $\sum_{i=1}^{3}\sum_{j=1}^{3}\langle\sigma_{i}\otimes\sigma_{j}\rangle^{2}\leq1$ according to Theorem 1, which results in $\sum_{i=1}^{3}\sum_{j=1}^{3}\delta^{2}(\sigma_{i}\otimes\sigma_{j})\geq 8 $, where $\delta^{2}(\sigma_{i}\otimes\sigma_{j})=\langle(\sigma_{i}\otimes\sigma_{j})^{2}\rangle-\langle \sigma_{i}\otimes\sigma_{j}\rangle^{2}$  is the variance of the observable $\sigma_{i}\otimes\sigma_{j}$. To prove the corollary 1, we introduce the principal components analysis (PCA) \cite{pea, hot, jol}, which is a mathematical procedure that transforms a number of possibly correlated variables into a number of uncorrelated variables called principal components. The first principal component accounts for as much of the variability in the data as possible, and each succeeding component accounts for as much of the remaining variability as possible. Similar to classical PCA, for the quantum covariance matrix $\gamma_{mm'}(\rho_{AB})$, the variances of principal components correspond to the eigenvalues of the covariance matrix, i.e., $\sum_{k=1}^{9}\lambda_{k}=\sum_{k=1}^{9}\delta^{2}P_{k}$, where $P_{k}$ is the principal component of the covariance matrix $\gamma_{mm'}(\rho_{AB})$, and $\sum_{k=1}^{9}\delta^{2}P_{k}=\sum_{i=1}^{3}\sum_{j=1}^{3}\delta^{2}(\sigma_{i}\otimes\sigma_{j})$, one has $\sum_{k=1}^{9}\lambda_{k}=\sum_{i=1}^{3}\sum_{j=1}^{3}\delta^{2}(\sigma_{i}\otimes\sigma_{j})$. So one attains $\sum_{k=1}^{9}\lambda_{k}\geq 8 $ for an unsteerable state. A detailed proof is provided in the Appendix A.

\section{Optimized steering criteria for symmetric two-qubit quantum systems}

Symmetry is another central concept in quantum theory \cite{gro}, which can be used to simplify the study of the entanglement sometimes \cite{voll, stoc, toth}. A bipartite quantum state $\rho$ is called symmetric if it is permutationally invariant, i.e., $F\rho F=\rho$, here $F=\sum_{ij}|ij\rangle\langle ji|$ is the flip operator. In the following we optimize the steering criterion for symmetric two-qubit quantum states.

\emph{Theorem 2.} If a given symmetric two-qubit quantum state is unsteerable from Alice to Bob (or Bob to Alice), the following inequality holds:
\begin{equation}
\sum\limits_{i=1}\limits^{3}\langle\sigma_{i}\otimes\sigma_{i}\rangle^{2}\leq1,
\end{equation}
where $\sigma_{i}$ ($i=1,2,3$) are Pauli operators.

\emph{proof.}
 For arbitrary symmetric two-qubit quantum state, one has $\langle\sigma_{i}\otimes\sigma_{j}\rangle=0$, where $i, j=1,2,3, i\neq j$. So \emph{Theorem 1} reduces to \emph{Theorem 2}.

\emph{Corollary 2.} If a given symmetric two-qubit quantum state $\rho_{AB}$ is unsteerable, the sum of the eigenvalues of the covariance matrix $\gamma_{mm'}(\rho_{AB})$ of the observables $\{O_{i}\}=\{\sigma_{i}\otimes\sigma_{i}\} (i=1,2,3)$ must satisfy
\begin{equation}\label{7}
\sum\limits_{k=1}\limits^{3}\lambda_{k}\geq 2,
\end{equation}
where $\lambda_{k}$ is the eigenvalue of the covariance matrix $\gamma_{mm'}(\rho_{AB})$. A brief proof of our theorem is specified below.

\emph{proof.} For a symmetric unsteerable state $\rho_{AB}$, one has $\sum_{i=1}^{3}\delta^{2}\langle\sigma_{i}\otimes\sigma_{i}\rangle^{2}\leq1$ from Eq.(9), which results in $\sum_{i=1}^{3}\delta^{2}(\sigma_{i}\otimes\sigma_{i})\geq 2 $.  For the quantum covariance matrix $\gamma_{mm'}(\rho_{AB})$, one has $\sum_{k=1}^{3}\lambda_{k}=\sum_{i=1}^{3}\delta^{2}(\sigma_{i}\otimes\sigma_{i})$ according to PCA. So one get $\sum_{k=1}^{3}\lambda_{k}\geq 2 $ for a symmetric unsteerable state .

\section{Illustrations of generic examples}

(i) \emph{Werner state.} Consider two-qubit Werner states \cite{wern}, which can be written as
 \begin{equation}\label{8}
 \rho_{W}=p|\psi^{+}\rangle\langle\psi^{+}|+(1-p)\mathbb{I}/4,
 \end{equation}
 where $|\psi^{+}\rangle=(1/\sqrt{2})(|00\rangle+|11\rangle)$ is Bell state and $\mathbb{I}$ is the identity, $0\leq p\leq1$. The Werner states are entangled iff $p>1/3$, steerable iff $p>1/2$ \cite{wis}, and Bell nonlocal if $p>1/\sqrt{2}$. According to symmetry of the Werner state and our \emph{Theorem 2}, we achieve $p>\sqrt{3}/3$ for successful steering under the Pauli measurements $\{\sigma_{1}, \sigma_{2}, \sigma_{3}\}$. Our results are in agreement with the results of Ref. \cite{zhe1, zhe2, zhen}, which implies that the nonlinear steering criterion is qualified for witnessing steering .

(ii) \emph{Bell diagonal states.} Suppose now that Alice and Bob share a  Bell diagonal state as follows:
\begin{equation}
 \rho_{bd}=\frac{1}{4}(\mathbb{I}+\sum_{i=1}^{3}c_{i}\sigma_{i}\otimes\sigma_{i}),
\end{equation}
 where $\sigma_{i}$ $(i=1,2,3)$ are Pauli operators and $|c_{i}|\leq1$ for $i=1,2,3$. According to \emph{Theorem 2}, we find that $\rho_{bd}$ are steerable if $\sum_{i}c_{i}^{2}>1$. In this case, the local uncertainty relations steering criterion can be written as $\sum_{i}\delta^{2}(\sigma_{i}^{B})-C^{2}(\sigma_{i}^{A},\sigma_{i}^{B})/\delta^{2}(\sigma_{i}^{A})>2$ \cite{zhen}, where $\delta^{2}(A)=\langle A^{2}\rangle-\langle A\rangle^{2}$ is the variance and $C(A,B)=\langle AB\rangle-\langle A\rangle\langle B\rangle$ is the covariance. The violation is $\sum_{i}c_{i}^{2}>1$ and the corresponding states are steerable. Likely for the linear criterion we have $|\sum_{i}\omega_{i}\langle\sigma_{i}^{A}\otimes\sigma_{i}^{B}\rangle|\geq\sqrt{3}$ with $\omega_{i}\in\{\pm1\}$ \cite{sau}, and the violation implies $|c_{1}\pm c_{2}\pm c_{3}|>\sqrt{3}$. For entropic criterion we have $\sum_{i}H(\sigma_{i}^{B}|\sigma_{i}^{A})>2$ \cite{sch2}, where $H(B|A)=\sum_{a}p(a|A)H(B|A=a)$ and $H(\cdot)$ denotes von Neumann entropy. The violation is $\sum_{i}(1+c_{i})log(1+c_{i})+(1-c_{i})log(1-c_{i})>2$. It can be checked that our criterion performs equivalently well as the local uncertainty relations steering criterion, which certifies more steerable states than the linear criterion and the entropic criterion (Fig.1).
\begin{figure}
\centering\includegraphics[width=0.6\columnwidth]{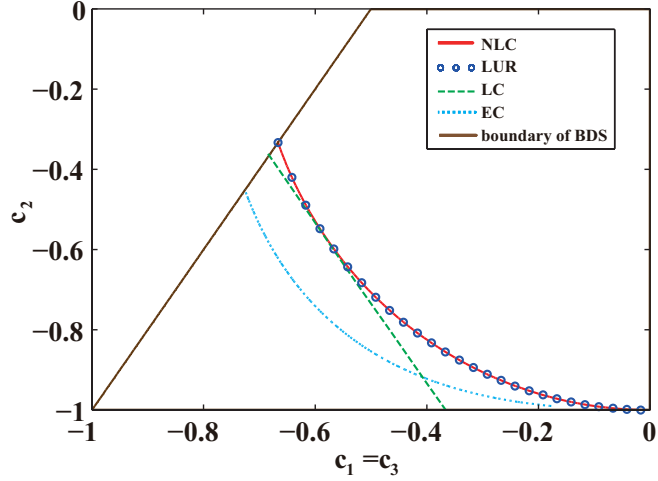}
\caption{The performances of different quantum steering criteria for the Bell diagonal states under the conditions $c_{1}=c_{3}$. The area inside the brown solid lines denotes Bell diagonal states (BDS). The red solid line, blue circled line, green dashed line, cyan dotted line are given by the nonlinear steering criterion (NLC),  local uncertainty relations criterion (LUR), linear criterion (LC), entropic criterion (EC), respectively. States in the left side of these lines are steerable.  It is clear that the NLC performs equivalently well as the LUR criterion, which certifies more steerable states than the LC and EC.}
\label{fig1}
\end{figure}

(iii) \emph{Asymmetric entangled states.} Let us consider Gisin states \cite{gis}, which can be expressed as
\begin{equation}\label{10}
\rho_{G}=p|\psi_{\theta}\rangle\langle\psi_{\theta}|+(1-p)\rho_{s},
\end{equation}
where $\psi_{\theta}=sin\theta|01\rangle+cos\theta|10\rangle$, $\rho_{s}=\frac{1}{2}|00\rangle\langle00|+\frac{1}{2}|11\rangle\langle11|$. In Fig.2, we show the performances of the nonlinear steering criterion (\emph{Theorem 1}), the local uncertainty relations steering criterion \cite{zhen}, the linear criterion \cite{sau} and the entropic criterion \cite{sch2} for the Gisin states. It follows from straightforward calculation that the nonlinear steering criterion certifies more steerable states than the linear criterion and entropic criterion.
\begin{figure}
\centering\includegraphics[width=0.6\columnwidth]{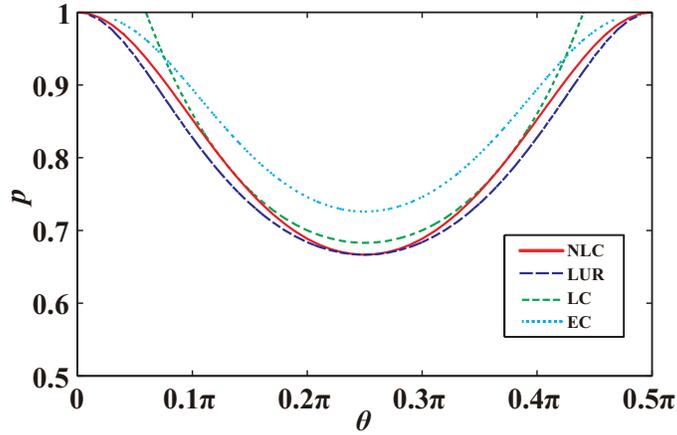}
\caption{The performances of different quantum steering criteria for the  Gisin states. The cyan dotted line, green dashed line, red solid line, blue dashed line are given by the EC, LC, NLC, LUR criterion, respectively. States above these lines are steerable.  It is clear that the NLC certifies more steerable states than the LC and EC.}
\label{fig1}
\end{figure}

\section{Conclusion}

In summary, we have proposed some nonlinear steering criteria applicable for arbitrary two-qubit quantum systems and optimized ones for symmetric quantum states. These criteria can be used to detect a wide range of steerable quantum states under Pauli measurements. Compared with the existing linear steering criterion and the entropic criterion, ours can certify more
steerable states without selecting measurement settings or correlation weights, which can also be used to verify entanglement
as all steerable quantum states are entangled.

\section*{Acknowledgments}
This work is supported by the National Natural Science Foundation of China (NSFC) under Grant No. 11947102, the Natural Science Foundation of Anhui Province under Grant Nos. 2008085MA16 and 2008085QA26, the Key
Program of West Anhui University under Grant No.WXZR201819,  the Research Fund for high-level talents of West Anhui University under Grant No.WGKQ202001004.

\appendix

\section{Proof of the equation $\sum_{k=1}^{9}\lambda_{k}=\sum_{i=1}^{3}\sum_{j=1}^{3}\delta^{2}(\sigma_{i}\otimes\sigma_{j})$}
In order to prove the Eq. $\sum_{k=1}^{9}\lambda_{k}=\sum_{i=1}^{3}\sum_{j=1}^{3}\delta^{2}(\sigma_{i}\otimes\sigma_{j})$, we extend  principal components analysis to quantum correlation matrix $\gamma_{mm^{'}}(\rho_{AB})$ of local observables $\{O_{m}\}=\{\sigma_{i}\otimes\sigma_{j}\} (i,j=1,2,3, m=3(i-1)+j)$. As in classical correlation analysis, the principal components on a matrix space can be expressed as
\begin{equation}
P_{j}=a_{1j}O_{1}+a_{2j}O_{2}+...+a_{9j}O_{9},
\end{equation}
where $j=1,2,...,9$. $\sum_{i}a_{ij}^{*}a_{ij}=1$, and $\sum_{i}a_{ij}^{*}a_{ik}=0$ for $j\neq k$.

To achieve the first  principal component, we use the Lagrange multiplier technique to find the maximum of a function. The Lagrangean function is defined as
\begin{eqnarray}
 L(a)=tr[\rho(a_{11}O_{1}+a_{21}O_{2}+...+a_{91}O_{9})^{2}]-\{tr[\rho(a_{11}O_{1}+a_{21}O_{2}+...+a_{91}O_{9})]\}^{2}\nonumber \\
     +\lambda_{1}(1-a_{11}^{2}-a_{21}^{2}-...-a_{91}^{2}), \ \ \ \ \ \ \ \ \ \ \ \ \ \ \ \ \ \ \ \ \ \ \ \ \ \ \ \ \ \  \ \ \ \ \ \ \ \ \ \ \ \ \ \ \ \ \ \ \ \ \ \ \ \ \ \ \ \ \ \ \ \ \ \ \ \ \ \ \ \ \ \ \ \ \ \ \ \ \
\end{eqnarray}
where $\lambda_{1}$ are the Lagrange multipliers. The necessary conditions for the maximum are
\begin{equation}
\frac{\partial L}{\partial a_{11}}=0; \frac{\partial L}{\partial a_{21}}=0;...; \frac{\partial L}{\partial a_{91}}=0.
\end{equation}
By using the properties of the trace, we obtain
\begin{eqnarray}
\frac{\partial L}{\partial a_{i1}}=2a_{i1}tr(\rho O_{i}^{2})-2a_{i1}[tr(\rho O_{i})]^{2}+\sum\limits_{k=1,...,9,k\neq i}a_{k1}[tr(\rho O_{i}O_{k})+tr(\rho O_{k}O_{i})]\nonumber\\
-\sum\limits_{k=1,...,9,k\neq i}a_{k1}[tr(\rho O_{i})tr(\rho O_{k})+tr(\rho O_{k})tr(\rho O_{i})]-2\lambda_{1} a_{i1}=0. \ \ \ \ \ \ \ \ \ \ \ \
\end{eqnarray}
By rearranging the above expression, we get
\begin{eqnarray}
a_{i1}[tr(\rho O_{i}^{2})-(tr(\rho O_{i}))^{2}]+\{\sum\limits_{k=1,...,9,k\neq i}a_{k1}[tr(\rho O_{i}O_{k})+tr(\rho O_{k}O_{i})]\}/2\nonumber\\
-\sum\limits_{k=1,...,9,k\neq i}a_{k1}tr(\rho O_{i})tr(\rho O_{k})=\lambda_{1} a_{i1}.\ \ \ \ \ \ \ \ \ \ \ \ \ \ \ \ \ \ \ \ \ \ \ \ \ \ \  \ \ \ \ \ \ \ \ \ \ \ \ \ \ \ \ \ \ \ \  \ \ \ \ \ \ \ \ \ \
\end{eqnarray}
For $i=1,...,9$, the following eigenvalue problem is obtained in compact form:
\begin{equation}
  \gamma \textbf{\emph{a}}_{1}=\lambda_{1} \textbf{\emph{a}}_{1},
\end{equation}
where $\textbf{\emph{a}}_{1}=(a_{11},a_{21},...,a_{91})'$,
$\gamma_{ij}=(\langle O_{i}O_{j}\rangle+\langle
O_{j}O_{i}\rangle)/2-\langle O_{i}\rangle\langle O_{j}\rangle$,
which is exactly the quantum covariance matrix as defined in Eq.(6). It shows that $\textbf{\emph{a}}_{1}$ should be chosen to be an eigenvector of
the covariance matrix $\gamma$, with eigenvalue $\lambda_{1}$. The
variance of  the first principal component is
\begin{equation}
 V(P_{1})=tr(\textbf{\emph{a}}_{1}^{\dagger}\gamma \textbf{\emph{a}}_{1})=\lambda_{1}.
\end{equation}
Therefore, in order to obtain the maximum of the variance, $\textbf{\emph{a}}_{1}$ should be chosen as the eigenvector corresponding to the largest eigenvalue $\lambda_{1}$ of the covariance matrix. Similarly, for the second  principal component, in order to obtain the second maximum of the variance,  $\textbf{\emph{a}}_{2}$ should be chosen as the eigenvector corresponding to the second largest eigenvalue $\lambda_{2}$ of the covariance matrix. This is fully consistent with the classical principal components analysis since the variances correspond to the eigenvalues of the covariance matrix.

For a arbitrary covariance matrix $\gamma_{ij}(\rho_{AB})$ of local observables $\{O_{m}\}=\{\sigma_{i}\otimes\sigma_{j}\} (i,j=1,2,3, m=3(i-1)+j)$, the variance of the observables $O_{m}$ can be analytically given as $\sum_{m=1}^{9}\delta^{2}(O_{m})=\sum_{i=1}^{N}\delta^{2}P_{i}$ due to the fact $\sum_{j}a_{ij}^{*}a_{ij}=1$. As $\sum_{i=1}^{N}\delta^{2}P_{i}=\sum_{i=1}^{N}\lambda_{i}$, one achieves $\sum_{m=1}^{9}\delta^{2}(O_{m})=\sum_{i=1}^{9}\lambda_{i}$.

\end{document}